\documentstyle[emulateapj,onecolfloat]{article}
\input epsf

\begin{document}

\twocolumn[
\title{Constraints on the long-range properties of gravity from weak
gravitational lensing}
\author{Martin White \& C.S. Kochanek}
\affil{ Harvard-Smithsonian Center for Astrophysics, Cambridge, MA 02138}
\authoremail{mwhite, ckochanek@cfa.harvard.edu}

\begin{abstract}
\noindent
\rightskip=0pt
Weak gravitational lensing provides a means of testing the long-range
properties of gravity.  Current measurements are consistent with standard
Newtonian gravity and inconsistent with substantial modifications on
Mpc scales.  The data allows long range gravity to deviate from a $1/r$
potential only on scales where standard cosmology would use normal
gravity but be dominated by dark matter.  
Thus, abnormal gravity theories must introduce two fine-tuning scales -- an
inner scale to explain flat rotation curves and an outer scale to force a
return to Newtonian gravity on large scales -- and these scales must
coincidently match the scales produced by the dark matter theory after
evolving the universe for 10 billion years starting from initial conditions
which are exquisitely determined from the cosmic microwave background.  
\end{abstract}
]

%\rightskip=0pt
\section{Introduction} \label{sec:intro}

Weak lensing of background galaxies by foreground large-scale structure
offers an opportunity to directly probe the mass distribution on large
scales over a wide range of redshifts.  As first pointed out by
Blandford et al.~(\cite{Blaetal91}) and Miralda-Escude~(\cite{Mir91}),
these effects are of order a few percent in adiabatic cold dark matter
models making their observation challenging but feasible.
Early predictions for the power spectrum of the shear and convergence were
made by Kaiser~(\cite{Kai92}) on the basis of linear perturbation theory.
Jain \& Seljak~(\cite{JaiSel97}) estimated the effect of
non-linearities in the density through analytic fitting formulae
(Peacock \& Dodds~\cite{PeaDod96}) and showed they substantially increase
the power in the convergence below the degree scale.  Because weak
lensing can measure the matter power spectrum without many of the
problems of approaches based on the distributions of galaxies or 
clusters (e.g. bias), it may ultimately provide as clean a cosmological 
probe as the microwave background.  Recently, several observational
groups have reported convincing evidence of the effect
(van Waerbeke et al.~\cite{Ludo};
 Bacon et al.~\cite{BRE};
 Kaiser et al.~\cite{KWL};
 Wittman et al.~\cite{WTKDB};
 Maoli et al.~\cite{Maoli};
 Rhodes et al.~\cite{RRG};
 van Waerbeke et al.~\cite{Ludo2})

All these theoretical and observational studies are primarily motivated by
standard theories of gravity and cosmology.  Despite the tremendous 
overall success of these theories, there has been a recent resurgence
of interest in non-standard theories of gravity, largely motivated by
the possibility that the standard paradigm has difficulty matching the
dynamical structure of galaxies (e.g. Flores \& Primack~\cite{Flores94};
Moore~\cite{Moore94}; Navarro \& Steinmetz~\cite{Navarro00}).
Most of these proposed modifications aim to make gravity a longer-ranged
force on scales comparable to the sizes of galaxies in order to explain
the flat rotation curves of galaxies on scales larger than the apparent
distribution of matter (e.g. Sellwood \& Kosowsky~\cite{Sellwood00}, 
Sanders~\cite{San98,San99,San00}; McGaugh~\cite{McGaugh99,McGaugh00},
but see Scott et al.~\cite{SWCP} and Aguirre et al.~\cite{ABFN} for an
opposite perspective).
As has been noted before (e.g.~Krisher~\cite{Kri}; Walker~\cite{Wal};
Bekenstein \& Sanders~\cite{BekSan}; Zhytnikov \& Nester~\cite{ZhyNes};
Edery~\cite{Ede}; Kinney \& Brisudova~\cite{Kinney01};
Uzan \& Bernardeau~\cite{UzaBer}; Mortlock \& Turner~\cite{MorTur})
any longer ranged gravitational force, if it also affects photons, should
have implications for gravitational lensing.
In particular it should profoundly affect the strength of weak lensing shears
on large scales.  Many of the above authors, however, consider gravitational
lensing by isolated objects.
To understand the lensing effects of modifying gravity on large scales it is
necessary to use the weak lensing formalism, summing over the contributions
{}from all density perturbations.

\section{The model} \label{sec:model}

We base our models on the discussion by Zhytnikov \& Nester~(\cite{ZhyNes})
of modified gravity theories within the context of linearized relativity
(see also Edery~\cite{Ede}).
This framework provides a relativistic gravity model which automatically
obeys the equivalence principle and within which definite calculations can be
made, while at the same time being as unrestrictive as possible.
Further discussion of the experimental foundations for the assumptions can
be found in Zhytnikov \& Nester~(\cite{ZhyNes}) and in
Weinberg~(\cite{Wei}), Misner, Thorne \& Wheeler~(\cite{MTW}) and
especially Will~(\cite{Wil}, \S\S2-3).

For any such model, the important change in the formalism for the propagation
of light through such a weak field metric is to change the Poisson equation
relating the density to the potential whose derivative is used to determine
the bend angle of photons.
The angular power spectrum of the convergence, $\kappa$, can be written as
an integral over the line-of-sight of the power spectrum of the density
fluctuations (Kaiser~\cite{Kai92}).  For sources at a distance $D_s$,
\begin{eqnarray}
  \ell(\ell+1)C_\ell/(2\pi) = {9\pi\over 4\ell}
    \left[\Omega_m H_0^2 D_s^2\right]^2 \int {d D\over D_s} t^3 (1-t)^2 
\nonumber \\
    \qquad\times
    \left[{\Delta_{\rm mass}^2(k=\ell/D,a)\over a^2}\right] 
f^2(k=\ell/D),
\label{eqn:semianalytic}
\end{eqnarray}
where $t\equiv D/D_s$, $\Delta_{\rm mass}^2(k)=k^3P(k)/(2\pi^2)$
is the contribution to the mass variance per logarithmic interval
physical wavenumber and $\ell(\ell+1)C_\ell/(2\pi)$ is the contribution to
$\kappa_{\rm rms}^2$  per logarithmic interval in angular wavenumber
(or equivalently multipole) $\ell$. 
The only change from the standard result is that the Poisson equation 
relating the potential to the density perturbations is modified from
$f(k)=1$ to a functional form determined by the Poisson equation of the
modified theory of gravity.  On small physical scales (large wavenumber
$k$), $f(k)=1$ is required to be consistent with the known properties
of gravity.

If the sources have a range of redshifts then one simply integrates the
above expression over the redshift distribution of the sources.  We shall
assume throughout that
\begin{equation}
  {dn\over dD} \propto D\exp[-(D/D_*)^4]
\end{equation}
and fix $D_*$ by the requirement that $\langle z_{\rm src}\rangle=1$.
In evaluating Eq.~(\ref{eqn:semianalytic}), we will use the method of
Peacock \& Dodds~(\cite{PeaDod96}) to compute the non-linear power spectrum
as a function of scale-factor.  Throughout we shall use the concordance
cosmology of Ostriker \& Steinhardt~(\cite{OstSte}) since it provides a
reasonable fit to recent CMB, weak lensing and large-scale structure data.
For this choice of parameters the lensing kernel peaks at $z\simeq 0.43$
at a (comoving) angular diameter distance of $1150h^{-1}\,$Mpc.

\begin{figure}
\begin{center}
\leavevmode
\epsfxsize=3.5in \epsfbox{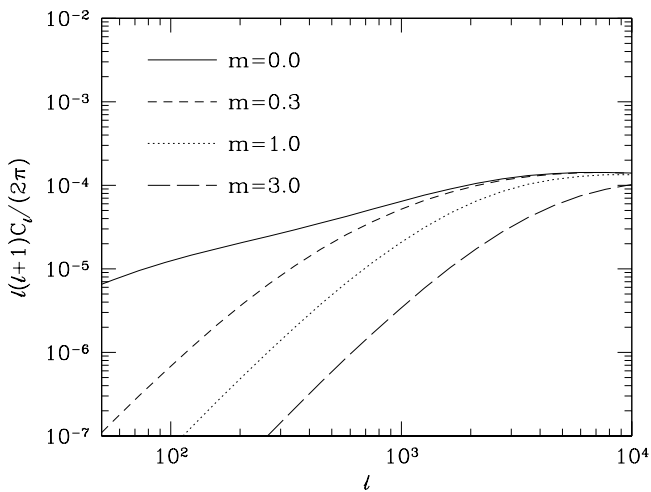}
\end{center}
\caption{\footnotesize%
The angular power spectrum, $\ell(\ell+1)C_\ell/(2\pi)$, vs.~multipole
moment $\ell$ for models with $\alpha=1.0$ and $m=0$, 0.3, 1.0 and
$3.0h\,{\rm Mpc}^{-1}$.  The sources are assumed to have
$\langle z_{\rm src}\rangle=1$.  Spectra for other values of $\alpha$ can
be roughly obtained by averaging the $m=0$ spectrum and the appropriate
$\alpha=1$ spectrum (plotted here) with the relevant weights.}
\label{fig:cl}
\end{figure}

In our calculation we only consider the propagation of rays through a known
density distribution, and we model that known density distribution using a
standard cosmological model viewed as a means to interpolate the evolution
of structure with redshift.  We do not attempt to self-consistently form
the observed structures using the modified gravitational potential\footnote{In
the model described below, a linear fluctuation analysis suggests that
long-wavelength modes would grow more slowly than the standard model would
predict.  Thus neglect of this effect is conservative if we start from an
initially scale-invariant spectrum.}.
If we assume that all theories must match the local density distribution,
the only consequence of this assumption is that the evolution of structure
implicit in Eq.~(\ref{eqn:semianalytic}) uses the standard growth rates
rather than those of the modified gravity.

Examining the effects of modified gravity simply becomes a question of
considering different structures for the function $f(k)$.   
In 4D, the metric, being symmetric, contains 10 functions.  The 4 constraints
of energy-momentum conservation reduce the number of free functions to 6.
These 6 free functions can be decomposed under rotations as 2 scalar
(density perturbations), 2 vector (vortical motions) and 2 tensor (gravity
wave) modes.
Within the linearized theory there are a number of propagating modes, which
have the form of Yukawa (exponential) potentials
\begin{equation}
  U(\vec{r};m) = G\int {\rho(\vec{r'}) d^3r'\over
               \left| \vec{r}-\vec{r}' \right| }
               e^{-m|\vec{r}-\vec{r}'|} \qquad .
\end{equation}
Under a variety of reasonable assumptions Zhytnikov \& Nester~(\cite{ZhyNes})
conclude that the most general metric describes forces mediated by massive and
massless scalar and tensor particles.
We follow Zhytnikov \& Nester~(\protect\cite{ZhyNes}) in neglecting the vector
modes, however we will allow arbitrary couplings for the scalar and tensor
modes.
%[in possible violation of the theorem of Jagannathan \&
%Singh~(\cite{JagSin})].
In general relativity in the weak field limit
\begin{eqnarray}
g_{00} &=& \left(-1 + 2U\right)\\
g_{ij} &=& \left(\phantom{-}1+2U\right)\delta_{ij}
\end{eqnarray}
where $U$ is the usual Newtonian potential.  The metric of
Zhytnikov \& Nester~(\cite{ZhyNes}) has the same form, but with Yukawa
potentials in addition to the Newtonian one.

\begin{figure}
\begin{center}
\leavevmode
\epsfxsize=3.5in \epsfbox{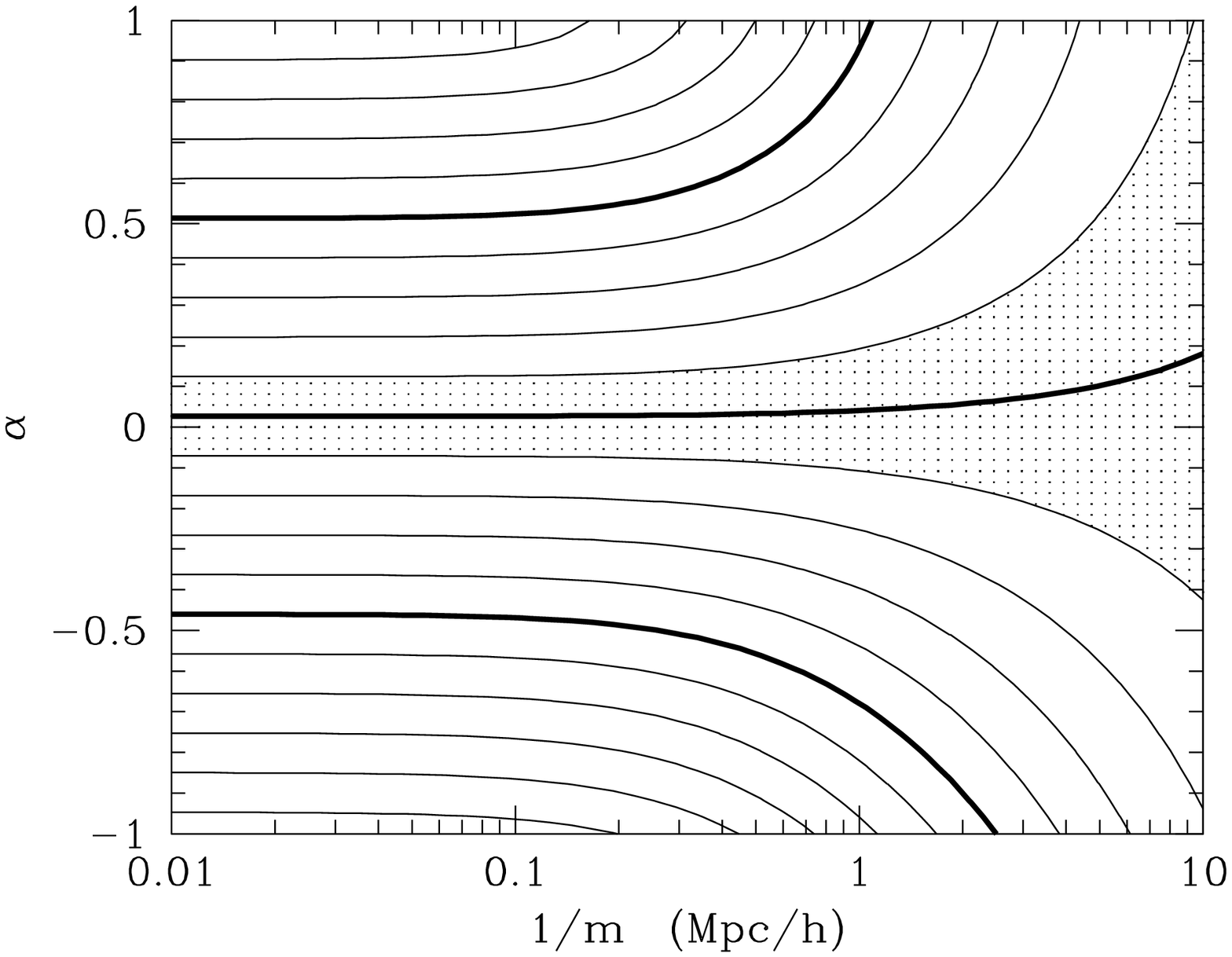}
\end{center}
\begin{center}
\leavevmode
\epsfxsize=3.5in \epsfbox{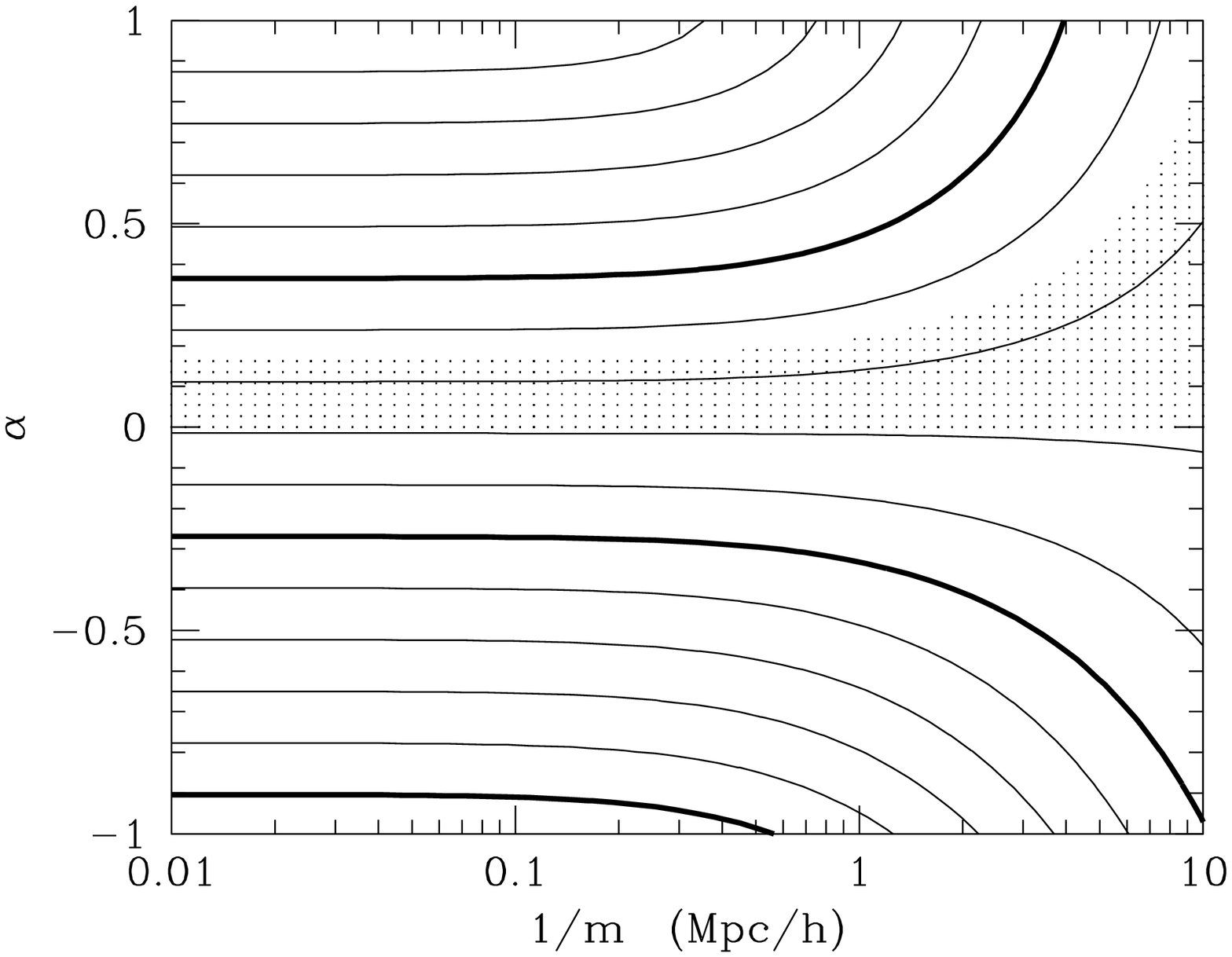}
\end{center}
\caption{\footnotesize%
The rms shear, smoothed with a $5'$ FWHM gaussian (top) or a $10'$ FWHM
gaussian (bottom), predicted for the ``concordance'' cosmology with
$\Omega_{\rm mat}=0.3$, $\Omega_\Lambda=0.7$, $h=0.67$ and $\sigma_8=0.9$
as a function of $\alpha$ and $m$ ($h$/Mpc).  Contours are spaced every
$0.001$ with bold contours indicating $0.005$ (top), $0.01$ and $0.015$
(bottom). The stippled regions are consistent (at $1\sigma$) with the 
van Waerbeke et al.~(\protect\cite{Ludo2}) measurements. }
\label{fig:sig5-10}
\end{figure}

For test particles with $v\ll c$ or fluids with $p\ll \rho c^2$ only the
time-time part of the metric is relevant, the contribution of the $g_{ij}$
terms being suppressed by ${\cal O}(v^2/c^2)$.
However, for light, the bend angle due to the potential is actually the
arithmetic mean of the coefficients in $g_{00}$ and $g_{ij}$.
Though the extra scalar and tensor modes can enter into the space-space and
time-time part of the metric differently, we shall consider the 1 parameter
family of models where these coefficients are equal.
As Kinney \& Brisudova~(\cite{Kinney01}) discuss,
%in suggesting that many features of MOND could be duplicated by a force law
%coupled only to baryon number,
the requirement that cluster mass estimates from galaxy dynamics,
pressure equilibrium of the X-ray gas and gravitational lensing agree means
that any modified gravity law must affect photon propagation in roughly the
same was as it affects particle orbits.  A modified gravity which
differentially affects particles and photons will almost always lead to a
discrepancy between these three cluster mass estimates.
 
\begin{figure}
\begin{center}
\leavevmode
\epsfxsize=3.5in \epsfbox{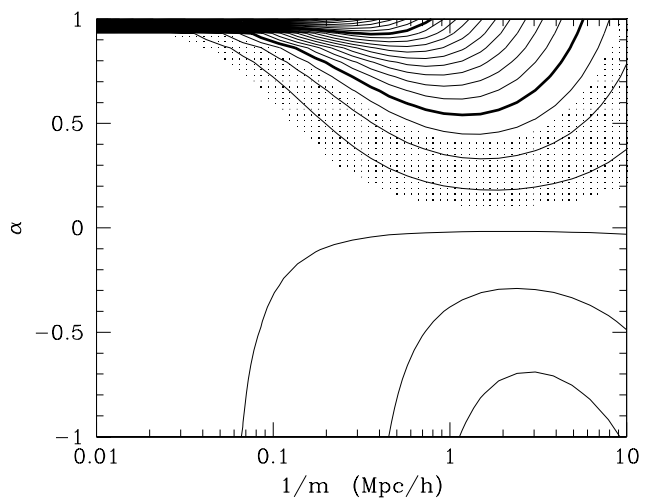}
\end{center}
\caption{\footnotesize%
The ratio of the rms shear on $5'$ and $10'$ scales for the same cosmology
as the previous figure.  Contours are spaced in steps of 0.05, increasing to
top left.  Thick contours are spaced every 0.5, starting at 1.5.
The stippled region is consistent (at $1\sigma$) with the
van Waerbeke et al.~(\protect\cite{Ludo2}) measurements.}
\label{fig:sigratio}
\end{figure}

Thus in our model, in the weak field limit, the propagation of light is
the same as in standard general relativity, except that the potential is
\begin{equation}
  U(\vec{r}) = (1-\alpha)U(\vec{r},0) + \alpha U(\vec{r},m) + \cdots
\end{equation}
where $\cdots$ represents possible other terms of the same form as
the second.  We shall further simplify our calculation by considering only 1
correction term in what follows.
In such a theory with one additional ``field'', the function appearing in
the estimate of the weak lensing power spectrum is
\begin{equation}
  f(k) = (1-\alpha) + \alpha\ {k^2\over k^2+m^2},
\end{equation}
where $\alpha=0$ for standard gravity and $\alpha \simeq -0.9$ and
$m^{-1} \sim 50\,$kpc in order to produce flat rotation curves without
dark matter (e.g.~Sanders~\cite{Sanders86}).
The corresponding potential for an object of mass $M$ simplifies to the
Newtonian result, $-GM/r$, on small scales where $mr\ll 1$, and has a
different effective coupling constant, $-GM(1-\alpha)/r$ on large scales,
$mr\gg 1$.

Fig.~\ref{fig:cl} shows the anisotropy spectrum predicted for a range of
models.  If we limit the range of gravity ($\alpha >0$) then the shear
fluctuations on large angular scales are suppressed, and if we extend the
range they are enhanced.
This should be a generic feature of any modification to the long-range force
law.  To obtain limits on the parameters in our model we calculated the rms
shear expected in Gaussian windows with FWHM of $5'$ and $10'$ as a function
of $\alpha$ and $m$ (Fig.~\ref{fig:sig5-10}).  These predictions are consistent
with the rms shear measured on these scales by
van Waerbeke et al.~(\cite{Ludo2}) only for models with parameters close
to those of standard gravity.
We can minimize the model dependence of the result by examining the ratio
of the power at $5'$ and $10'$, as this largely removes any dependence of
the result on the matter density and the normalization of the power spectrum.
In Fig.~\ref{fig:sigratio}, we see that the data are consistent with standard 
gravity and a broad range of alternate theories.  These theories
are acceptable because our alternate gravity model has a $1/r$ potential 
on large scales so that when the $5'$ scale corresponds to a physical
scale larger than $m^{-1}$, the change in the coupling constant $\alpha$
is degenerate with a change in the enclosed mass. For sources with a 
mean redshift of unity, the $5'$ scale corresponds to a length scale
at the peak of the lensing kernel of approximately $1h^{-1}\,$Mpc.

\begin{figure}
\begin{center}
\leavevmode
\epsfxsize=3.5in \epsfbox{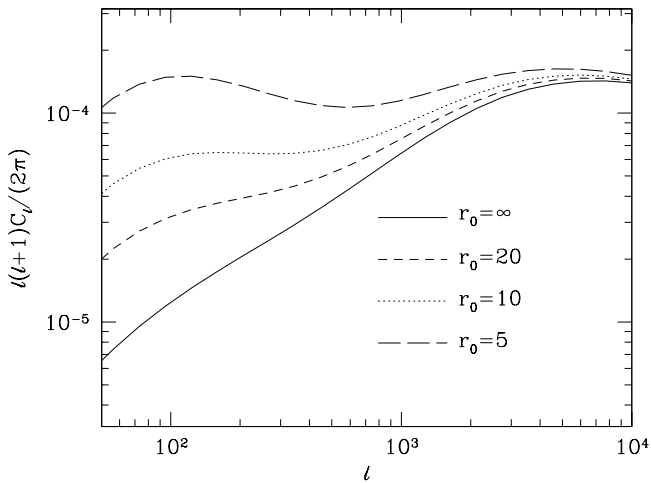}
\end{center}
\caption{\footnotesize%
The angular power spectrum, $\ell(\ell+1)C_\ell/(2\pi)$, vs.~multipole moment
$\ell$ for our second model with $m^{-1}=50h^{-1}\,$Mpc and $r_0=\infty$, $20$,
$10$ and $5h^{-1}\,$Mpc.  As $r_0\rightarrow\infty$ the model becomes standard
gravity.  Note the change in shape and the enormous enhancement in the power on
large scales.}
\label{fig:cl2}
\end{figure}

Theories which do not return to a $1/r$ potential on large scales are
relatively easy to rule out (see Walker~\cite{Wal}).  Assuming that the
bend angle of light remains proportional to the gradient of the projected
gravitational potential, such theories predict that random lines of sight
would be highly sheared and (de)magnified\footnote{For example, for a $\log r$
potential and a Poisson distribution of lenses the convergence, $\kappa$, of a
source at $D_s$ (assumed to be much larger than the scale, $r_0$, beyond which
gravity is $\log r$) is
$\kappa\simeq \pi\alpha_0\left(n r_0^2 D_s\right)\gg 1$ for any reasonable
source density $n$.}
in contradiction with observations.
This problem can be traced to the lack of degeneracy between renormalizing the
mass and adjusting the coupling constant.
For example, ignoring the Kinney \& Brisudova~(\cite{Kinney01}) {\it ansatz\/}
for permissible forms of alternate gravity, we could use the force law
\begin{equation}
    -\phi'(r)/GM = - { 1 \over r^2 } - { \exp(-m r) \over r r_0 } 
\end{equation}
which is Keplerian for $r\ll r_0$ and $r\gg m^{-1}$ but is a $1/r$ force law,
producing a flat rotation curve, in between.
The potential corresponding to this force law is
\begin{equation}
  \phi/GM= -1/r + {\rm Ei}[-mr]
\end{equation}
where ${\rm Ei}[x]$ is the exponential integral.
The corresponding kernel for the weak lensing integral is
\begin{equation}
  f(k) = 1 - { k m - (k^2+m^2) \tan^{-1}(k/m) \over r_0 k (k^2+m^2) }.
\end{equation}
Figure~\ref{fig:cl2} shows the angular power spectrum in this model for a
range of scales $r_0$ and a large outer cutoff $m^{-1}=50h^{-1}\,$Mpc.  
Compared to normal gravity, the modified theories have enormously enhanced
large scale power and very different shapes.

\section{Discussion} \label{sec:discussion}

Current modified gravity theories tuned to explain the rotation curves of
galaxies work in a standard cosmology because we measure rotation curves only
where there are baryons.  We can see that the rotation curve is flat out to
the limit where there are no more baryons to measure, but we cannot see that
it is Keplerian as we approach the edge of the more extended dark matter
distribution.
If we could continue to trace rotation curves on larger scales we would see
a growing difference between standard cosmological models and theories using
modified gravitational physics.  

Weak lensing allows us to do this experiment, although on such large scales
we must sum over the contributions of all of the mass rather than consider the
rotation curves of discreet objects.  As we would expect qualitatively,
increasing the strength of the gravitational field at long ranges predicts
stronger weak lensing signals on large scales than standard cosmological
models.  Current measurements of the rms shear on scales of $5'$-$10'$ 
rule out the theories we consider in the parameter ranges where they 
could explain rotation curves without dark matter unless the deviation from
normal gravity is limited to a restricted range of spatial scales from
$10h^{-1}\,{\rm kpc}\la r\la 1h^{-1}\,$Mpc.
On larger scales the models must return to the $r^{-2}$ force law of normal
gravity in order to be consistent with measurements.

In standard cosmological models, once we postulate the existence of dark 
matter, the inner and outer scales appear naturally.  On small scales the
cooling of the baryons concentrates the baryons relative to the dark matter
and renders them luminous and detectable.  Thus, normal matter combined with
normal gravity naturally explain dynamics on scales $\la 10h^{-1}\,$kpc.
On intermediate scales, dark matter provides an additional source of density,
which can be interpreted as an abnormal gravitational theory using only the
visible baryons as sources.  On large scales the universe returns to
homogeneity, and the special properties of the $1/r^2$ force law make the weak
lensing power slowly diminish on large scales.  Abnormal, longer ranged
theories lose the cancellation properties of the $1/r^2$ force law on large
scales, despite the increasing homogeneity of the density on these scales,
leading to enormous enhancements in the strength of the weak lensing shear.
Such strong shears are in gross disagreement with even the first generation
of weak lensing measurements on these scales
(van Waerbeke et al.~\cite{Ludo};
 Bacon et al.~\cite{BRE};
 Kaiser et al.~\cite{KWL};
 Wittman et al.~\cite{WTKDB};
 Maoli et al.~\cite{Maoli};
 Rhodes et al.~\cite{RRG};
 van Waerbeke et al.~\cite{Ludo2}).
Thus, abnormal gravity theories must introduce two fine-tuning scales -- an
inner scale to explain flat rotation curves and an outer scale to force a
return to Newtonian gravity on large scales -- and these scales must
coincidently match the scales produced by the dark matter theory after
evolving the universe for 10 billion years starting from initial conditions
which are exquisitely determined from the cosmic microwave background.  

Finally, although we lack a formalism for estimating weak lensing in
non-potential theories such as MOND, Mortlock \& Turner~(\cite{MorTur})
have emphasized that weak lensing results should be generic, as it requires
only that photons and particles have similar responses to gravitational fields.
This similarity of behavior is observed on the relevant scales (Mpc) through
the near equivalence of weak lensing, dynamical, and X-ray determinations of
cluster masses (Kinney \& Brisudova~\cite{Kinney01}).  

\bigskip
\acknowledgments  
We would like to thank L.V. van Waerbeke for providing more details of their
VIRMOS survey results.
M.W. was supported by NSF-9802362 and a Sloan Fellowship.
C.S.K. was supported by the Smithsonian Institution and NASA grants NAG5-8831
and NAG5-9265.

\end{document}